\documentclass[epj]{svjour}
\usepackage{graphicx}

\def\sech{\mbox{sech}}

\begin{document}
\title{Multi-component structure of nonlinear excitations\\
in systems with length-scale competition}
\author{Peter L. Christiansen \inst{1} 
\thanks{\emph{e-mail:} plc@imm.dtu.dk}
\and Yuri B. Gaididei \inst{2} 
\and Franz G. Mertens \inst{3} 
\and Serge F. Mingaleev \inst{2}
}
\institute{
Department of Mathematical Modelling,
The Technical University \\
of Denmark, DK-2800 Lyngby, Denmark 
\and 
Bogolyubov Institute for Theoretical Physics,
03143 Kiev, Ukraine
\and 
Physikalisches Institut, Universit{\"a}t
Bayreuth, D-95440 Bayreuth, Germany}
\date{\today}
\abstract{
We investigate the properties of nonlinear excitations in
different types of soliton bearing systems with 
{\em long-range dispersive interaction}. We show that
length-scale competition in such systems universally results
in a multi-component structure of nonlinear excitations 
and can lead to a {\em new type of multistability}: 
coexistence of different nonlinear excitations at the same 
value of the {\em spectral parameter} (i.e., velocity in 
the case of anharmonic lattices or frequency in 
nonlinear Schr{\"o}dinger models).
} 
\authorrunning{Christiansen \and Gaididei 
\and Mertens \and Mingaleev}
\titlerunning{Multi-component structure of nonlinear 
excitations} 
\maketitle
%

\section{Introduction}

Physical systems with competition between discreteness,
nonlinearity and dispersion are abundant in Nature.
Examples are nonlinear charge and energy transport in
biological macromolecules, elastic energy transfer in
anharmonic chains, charge transport in hydrogen-bonded
systems, coupled optical fibers, nonlinear photonic crystals,
arrays of coupled Josephson junctions. Determination of
their dynamical properties is of importance because of
their wide applicability in various physical problems.
In nonlinear systems with short-ranged (nearest-neighbor)
dispersive interactions there are basically two types of
nonlinear excitations. In the large dispersion
(weak nonlinearity) limit when the discreteness of the
lattice plays no essential role, the balance between
nonlinearity and dispersion provides the existence of
low-energy soliton-like excitations. The width
of these excitations is large comparing with the lattice
spacing, they freely propagate along the system without
energy loss, their collisions are almost elastic. When the
nonlinearity is strong (dispersion is weak) new nonlinear
excitations, intrinsically localized modes, also called
discrete breathers, may appear 
(see {\it e.g.} Ref. \cite{flach} for a review). 
Their width is comparable with the lattice
spacing, they are long-lived and robust.

The phenomenon of multistability, implying the coexistence
of several stationary localized states having the same
energy, appears in different contexts
\cite{lst,mal,kladko,ngfm,gfnm,mgm,gmck,gmcr,mks,jgcr,jagcr,gcrj,kov}.
For the one-dimensional discrete nonlinear Schr{\"o}dinger
(NLS) equ\-ation with nearest-neighbor interactions (NNI) and 
arbitrary power of nonlinearity,
multistability occurs when the power exceeds some critical
value \cite{lst,mal,kladko}.

Another mechanism of such multistability comes out for the 
systems with long-range dispersive interactions, where a 
new characteristic length scale appears: the 
radius of the dispersive interaction. As it was shown in
Refs. \cite{ngfm,gfnm,mgm,gmck,gmcr,mks} 
the appearance of the new characteristic length scale brings
into coexistence two stable stationary excitations
one of which is a broad, continuum-like soliton and another
is a narrow discrete excitation. This is a generic property
of nonlinear systems with nonlocal dispersive interactions.
The bistability occurs in such different systems as
anharmonic chains with nonlocal interatomic interaction
\cite{ngfm,gfnm,mgm}, NLS models 
with long-range coupling \cite{gmck,gmcr}, and nonlinear
waveguides in photonic crystals \cite{mks}. 
In all these systems the bistability means that there is 
an interval of the energy $H$ where two stable nonlinear
excitations (one of which is broad and another is 
intrinsically localized) coexist for each value of $H$. 
The greater the radius of the long-range interactions is, 
the more pronounced is the bistability phenomenon. 
A general mechanism for obtaining a controlled
switching between bistable localized excitations was
proposed in Refs. \cite{jgcr,jagcr}. Importantly, very similar
types of excitations may appear in periodic nonlinear
dielectric superlattices (two-dimensional nonlinear
Kronig-Penney model \cite{gcrj}) where the nonlocality
arises as a result of a reduction of the original
two-dimensional problem to an effective one-dimensional
description. Bistability in the spectrum of nonlinear
localized excitations in magnets with a weak exchange
interaction and strong magnetic anisotropy was found
in Ref. \cite{kov}.

Quite recently coexistence of different nonlinear 
excitations at {\em the same value of the soliton velocity} 
has been shown to take place in anharmonic chains with 
{\em ultra-long-range} interatomic interactions 
\cite{mgm00}. Specifically, it was shown that
three different types of pulse solitons can coexist for 
each value of the pulse velocity in a certain velocity 
interval: one type is unstable but the two others
can be stable. It should be emphasized the following 
intriguing aspect of the problem: in contrast 
to the previously discussed cases, now the bistability 
takes place at the same value of a {\em spectral parameter} 
(the soliton velocity in this system is mathematically 
the spectral parameter of a nonlinear eigenvalue problem). 
Therefore, the system being discussed provides us with the 
example of a {\em new type of multistability} whose physical
consequences can be quite different from the multistability 
discussed above. Indeed, if such a new type of multistability
can exist, for instance, for nonlinear waveguides 
in photonic crystals \cite{mks}, then two different
nonlinear modes with the same frequency of the light 
will coexist in a waveguide, which is practically more 
important than to have coexistence of nonlinear modes 
with the same energies but different frequencies. 

The aim of the present paper is to demonstrate that the new 
type of multistability is a fundamental property of
nonlinear systems with ultra-long-range dispersion. 
To this end we consider two additional (in comparison with 
Ref. \cite{mgm00}) models: a discrete 
NLS equation with nonlocal dispersion and an anharmonic 
lattice with screened Coulomb interaction. 
We show that in both cases an {\em ultra-long-range} 
dispersive interaction results in a 
{\em multi-valued dependence} of the energy
of stationary states on a spectral parameter (frequency in
the case of the NLS equation and velocity in the case
of the anharmonic lattice). 

The outline of the paper is the following. In Sec. 2 we
present the results of numerical simulations of stationary
states in the discrete NLS model with a long-range
dispersive interaction and in the anharmonic chain with a 
long-range dispersive interaction and cubic
nearest-neighbor interaction. We discuss the appearance of
a multi-valued dependence of the energy of the systems. 
In Sec. 3 we develop an
analytical approach to the problem. We use a quasicontinuum
approach which preserves both length scales that enter
our problem. We show that the low-energy excitation can be
considered as a compound state: a short-ranged breather-like
component and a long-ranged component.
Section 4 presents the concluding discussion.

\section{Models and numerical results}

We consider in parallel two nonlinear models which can bear
pulse-like excitations: the Boussinesq equation model which
describes the properties of anharmonic molecular chains and
the nonlinear Schr{\"o}dinger equation model which
represents generic properties of a system of nonlinear
oscillators.

\subsection{Anharmonic chains with ultra-long-range
dispersive interactions}

Let us consider a chain of equally spaced particles of unit
mass whose displacements from equilibrium are $u_n(t)$ and the
equilibrium spacings are unity. The Hamiltonian of the system
is given by
\begin{eqnarray}
\label{sys:hamil}
H= \sum_n \biggl\{ \frac{1}{2} \Bigl( \frac{du_n}{dt} \Bigr)^2
+ V(u_{n+1}-u_n) \nonumber \\
+ \frac{1}{2} \sum_{m>n} J_{m-n} (u_m-u_n)^2 \biggr\} \; ,
\end{eqnarray}
with the anharmonic interaction
\begin{eqnarray}
V(w)= w^2/2-w^3/3
\end{eqnarray}
between nearest neighbors and the harmonic long-range
interaction (LRI)
\begin{eqnarray}
\label{disp}
J_{m-n}(\alpha,s)= J(\alpha,s) \,
\frac{e^{-\alpha |m-n|}}{|m-n|^{s}}
\end{eqnarray}
between all particles of the chain.
Here
\begin{eqnarray}
\label{coeff}
J(\alpha,s)= J/F(e^{-\alpha},s) \; ,
\end{eqnarray}
where
\begin{equation}
\label{eq19}
F(z, s)=\sum_{n=1}^{\infty}(z^n/n^s)
\end{equation}
is the so-called Joncqi\`{e}re's function which properties
are described in Ref. \cite{htf53}.
$J$ characterizes the intensity of the LRI's whereas
$\alpha$ and $s$ determine their inverse radius. The
coefficient $J(\alpha,s)$ is chosen in the form
(\ref{coeff}) to have the sum $\sum_n\,J_{n}(\alpha,s)$
independent of $\alpha$ and $s$. The parameters $\alpha$
and $s$ are introduced to cover different physical
situations from the limit of nearest-neighbor interactions
($\alpha \gg 1$ or $s \gg 1$) to the limit of ultra-long-range
interactions ($\alpha \ll 1$ and $s \leq 3$).
The Hamiltonian (\ref{sys:hamil}) generates equations of
motion of the form
\begin{eqnarray}
\label{sys:eq-wn}
\frac{d^2 w_n}{dt^2} &+& 2 W(w_{n}) - W(w_{n+1})
- W(w_{n-1}) \nonumber \\
&+& \sum_{m \neq n} J_{m,n} (w_n-w_m)=0 \; ,
\end{eqnarray}
where $w_n=u_{n+1}-u_n$ are relative displacements and
$W(w) \equiv d V(w)/dw = w-w^2\,$.
We use the quasicontinuum approach, regarding $n$ as a
continuous variable:
$n \rightarrow x, \, w_n(t) \rightarrow w(x,t)$,
and using the operator identities
\begin{eqnarray}
\label{qcont}
w(x+m,t) &=& \, e^{m\partial_x} w(x,t) \; , \nonumber \\
Q(\alpha, s, \partial_x) &\equiv& 
\sum_{m=1}^{\infty} \frac{e^{-\alpha m}}{m^s} \,
\frac{1-\cosh m\partial_x}{F \left(e^{-\alpha}, s \right)}
\nonumber \\
&=& 1 - \frac{F \left(e^{-\alpha-\partial_x}, s \right) +
F\left(e^{-\alpha+\partial_x}, s \right)}{2 \,
F\left(e^{-\alpha}, s \right)} \; .
\end{eqnarray}
In this approach the equation of motion (\ref{sys:eq-wn})
can be written in the pseudo-differential form
\begin{equation}
\label{sys:eq-wx}
[\partial^2_t - J Q(\alpha, s, \partial_x)] w(x,t)
- 4 \sinh^2 \biggl( \frac{\partial_x}{2} \biggr)
W(w)=0 \; ,
\end{equation}
where $\partial_x$ and $\partial_t$ are the derivatives with
respect to $x$ and $t$, respectively. Eq. (\ref{sys:eq-wx})
can be called {\em non-local Boussinesq equation}.

\begin{figure}
\includegraphics[width=0.44\textwidth,clip]{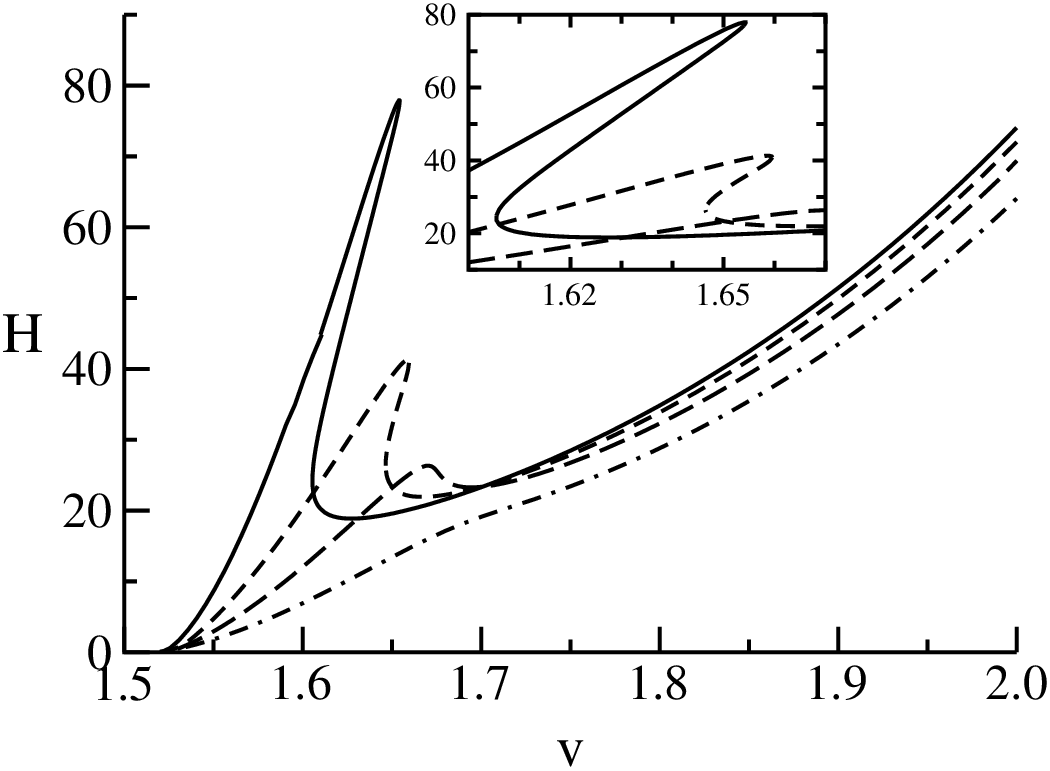}
\caption{Energy of the pulse solitons versus velocity
in the anharmonic lattice with Kac-Baker 
($s=0$) long-range
interaction found numerically for different values of
$\alpha$ and $J$ (the value of $J$ was chosen to
get constant $c=1.515$):
$\alpha=0.05$ and $J=0.0016$ (full line);
$\alpha=0.1$ and $J=0.0062$ (dashed line);
$\alpha=0.17$ and $J=0.0172$ (long-dashed line);
$\alpha=0.3$ and $J=0.05$ (dot-dashed line).}
\label{fig:energy:s=0}
\end{figure}

\begin{figure}
\includegraphics[width=0.44\textwidth,clip]{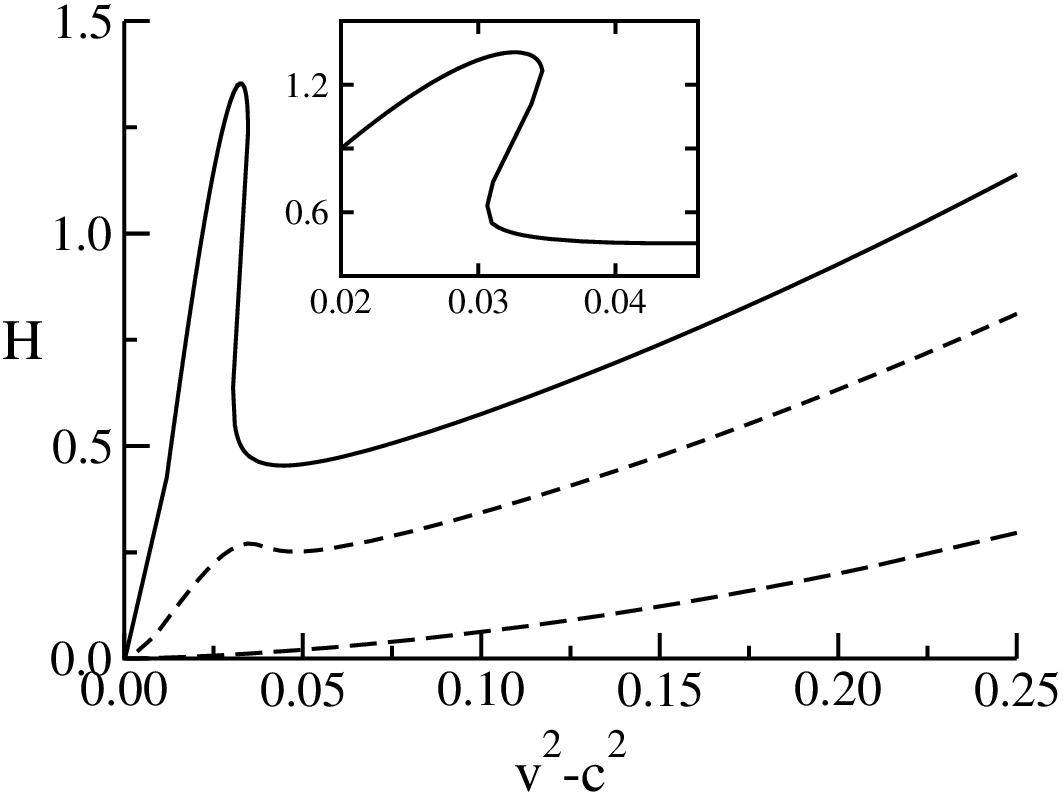}
\caption{Energy of the pulse solitons versus velocity
in the anharmonic lattice with screened Coulomb 
($s=3$) interaction
for $J=0.05$ and different values of $\alpha$:
$\alpha=0.001$ (full line), $\alpha=0.005$ (dashed line),
and $\alpha=\infty$ (long-dashed line).}
\label{fig:energy:s=3}
\end{figure}

\begin{figure}
\includegraphics[width=0.44\textwidth,clip]{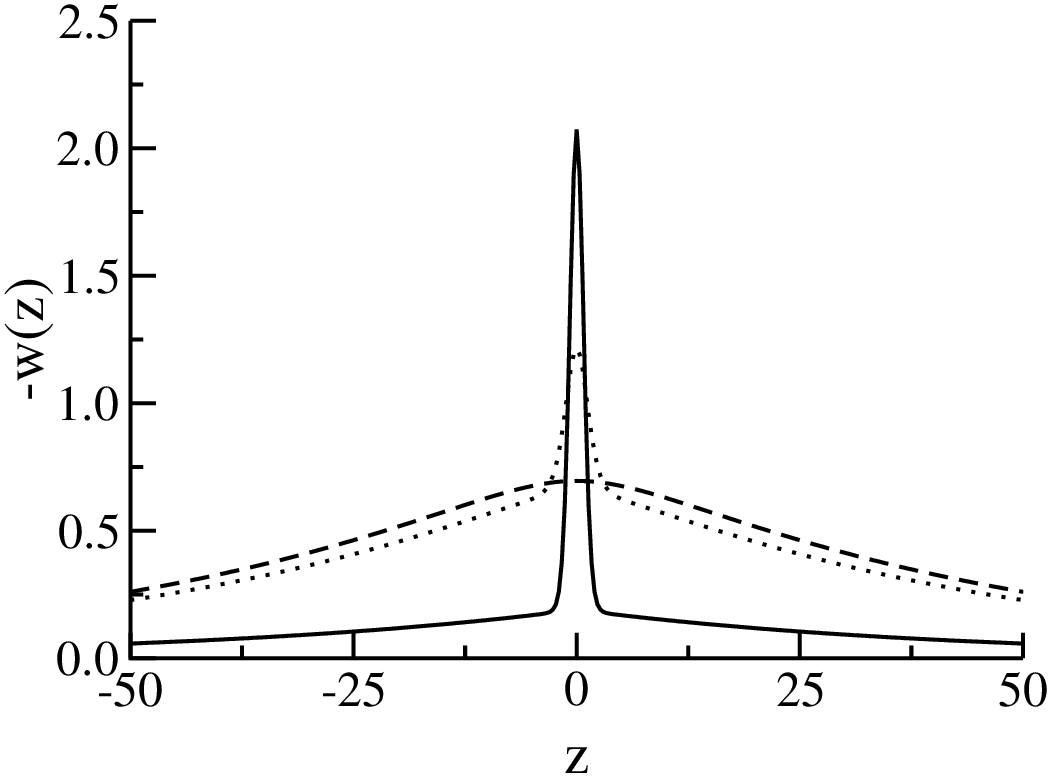}
\caption{Shapes of the low-energy (full line),
unstable intermediate-energy (dotted line), and high-energy
(dashed line) pulse solitons which coexist at the
same velocity $v=1.64$ in the anharmonic lattice with
Kac-Baker ($s=0$) 
long-range interaction for $\alpha=0.05$ and
$J=0.0016$ (see Fig. \protect\ref{fig:energy:s=0}).}
\label{fig:shape:s=0}
\end{figure}

\begin{figure}
\includegraphics[width=0.44\textwidth,clip]{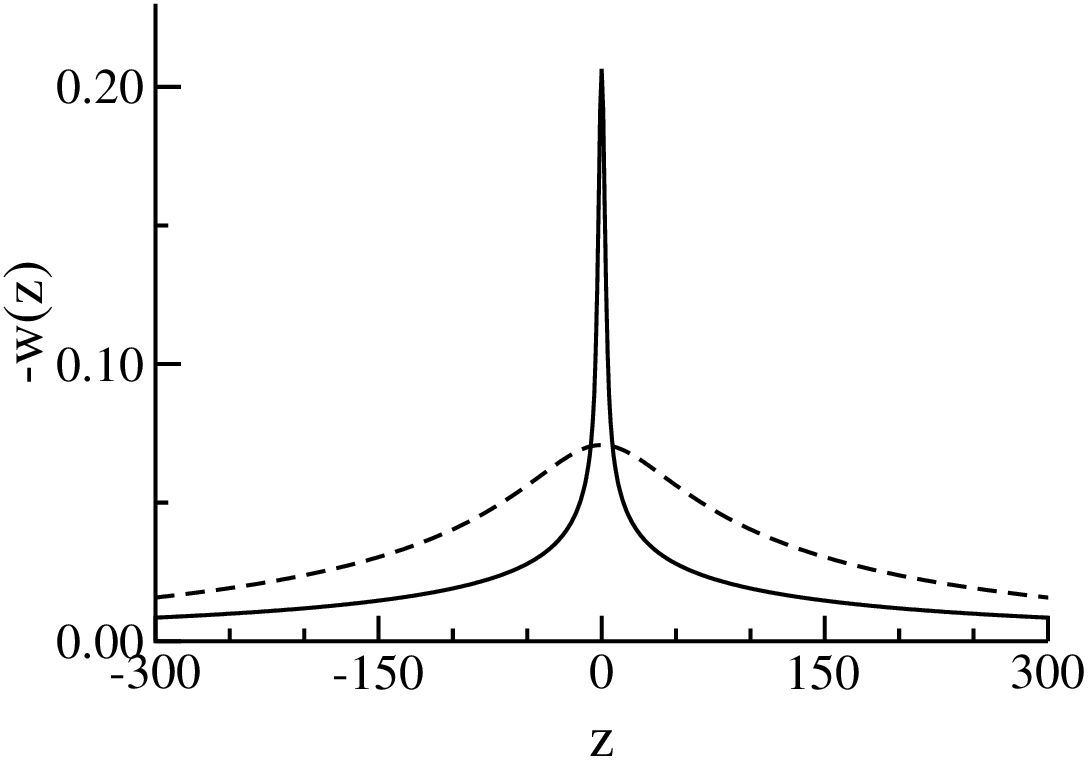}
\caption{Shapes of the low-energy (full line) and
high-energy (dashed line) pulse solitons which coexist
at the same velocity $v^2-c^2=0.031$ in the anharmonic
lattice with screened Coulomb 
($s=3$) interaction for $J=0.05$
and $\alpha=0.001$ (see Fig. \protect\ref{fig:energy:s=3}).}
\label{fig:shape:s=3}
\end{figure}

The energy (\ref{sys:hamil}) together with the momentum
\begin{eqnarray}
\label{momentum}
P=\int_{-\infty}^{\infty}
\Bigl( \frac{\partial u}{\partial t} \Bigr)
\Bigl( \frac{\partial u}{\partial x} \Bigr) \, dx
\end{eqnarray}
are conserved quantities.
We are interested in the stationary soliton solutions
$w(x,t) \equiv w(x-vt)$ propagating with velocity $v$.
Substituting $z=x-vt$ we can write Eq. (\ref{sys:eq-wx})
in the form
\begin{equation}
\label{sys:eq-st}
[v^2 \partial^2_z - J Q(\alpha, s, \partial_z)] w(z)
- 4 \sinh^2 \Bigl( \frac{\partial_z}{2} \Bigr)
W(w)=0 \; .
\end{equation}
In this way we reduce our problem to a nonlinear eigenvalue
problem with $v$ being a spectral parameter.

We consider two physically important cases: screened Coulomb
interactions ($s=3$) and the Kac-Baker LRI's ($s=0$).
The speed of sound $c$ (which is an upper limit of the group
velocity of linear waves), determined by the expression
$c^2=1 + J F\left(e^{-\alpha}, s-2 \right) /
F\left(e^{-\alpha}, s \right)$,
grows indefinitely in both cases as $\alpha$ decreases.
Using the Green's function method \cite{mgm} one can show
that spatially localized stationary  solutions (solitons)
exist only for supersonic velocities $v>c$.

We have numerically integrated Eq. (\ref{sys:eq-st}) using
the method developed in Ref. \cite{mgm}. Figures
\ref{fig:energy:s=0} and \ref{fig:energy:s=3} show
the dependence of the energy $H(v)$ (the dependence $P(v)$
has the same form) of the soliton solutions of
Eq. (\ref{sys:eq-st}) obtained for the Kac-Baker interaction
($s=0$) and screened Coulomb interaction ($s=3$),
respectively. It is seen that the behavior in both cases is
qualitatively the same. The soliton energy grows
{\em monotonically} with the velocity in the case of
large $\alpha$ (see, e.g., $\alpha=0.3$ in
Fig. \ref{fig:energy:s=0}).
In this case the soliton properties are qualitatively the
same as in the limit of nearest-neighbor interactions
(NNI's) for which Eq. (\ref{sys:eq-wx})
reduces to the Boussinesq equation. It is well known that
this equation has a sech-shaped soliton solution
$w(z)= -1.5 \, (v^2-c^2)/\cosh^2 (\sigma z)$, where
$\sigma=\sqrt{3(v^2-c^2)}$ is the inverse soliton width.
As indicated above, the  energy $H \sim (v^2-c^2)^{3/2}$ of
these solitons is a monotonic function of the velocity,
which means that there is only one soliton solution for each
given value of energy or velocity.

The situation changes when $\alpha < \alpha_{cr}$ (e.g., 
for the Kac-Baker interaction the critical value of 
inverse radius $\alpha_{cr}(J)$ is determined by Eq. 
(9) in Ref. \cite{mgm00}). 
In this case, as was shown first in Ref.
\cite{gfnm}, the dependence $H(v)$ becomes {\em non-monotonic}
(of $N$-shape, with a local maximum and a local minimum),
so that there exist two branches of {\em stable} supersonic
solitons: low-velocity and high-velocity solitons. These two
soliton branches are separated by a {\em gap} (an interval
of velocities where $H(v)$ decreases) with {\em unstable}
soliton states. It is interesting that because of
the non-monotonic dependence of $H(v)$ there exist
low-velocity and high-velocity solitons of equal energy.
Further increasing of the radius of the dispersive
interaction (decreasing $\alpha$) leads to a $Z$-shape
dependence of $H(v)$ (see Figures \ref{fig:energy:s=0} and
\ref{fig:energy:s=3}). In this case the gap which separated
two above-mentioned  soliton branches collapses; instead,
there appears an interval of velocities where three soliton
solutions with different energies and shapes (see Figures
\ref{fig:shape:s=0} and \ref{fig:shape:s=3}) coexist for
each value of velocity. This phenomenon was first
observed in Ref. \cite{mgm00} for the Kac-Baker dispersive
interaction. Here we show that the same phenomenon takes
place also in the case of the screened Coulomb interaction,
but the radius of the interaction should be much larger:
{\em e.g.}, for the Coulomb dispersive interaction with
the intensity $J=0.05$ the gap vanishes for
$\alpha<0.0015$ while for the Kac-Baker interaction
with the same intensity $J$ it happens for $\alpha<0.15$.

\begin{figure}
\includegraphics[width=0.44\textwidth,clip]{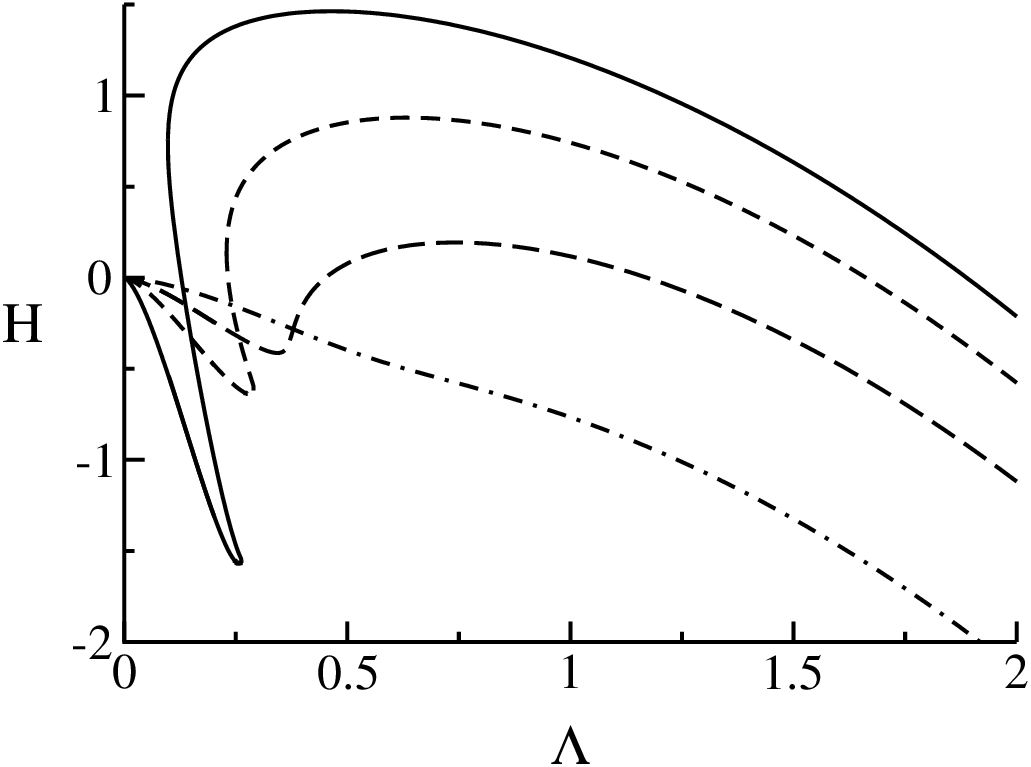}
\caption{Energy $H$ of the
stationary solitons versus nonlinear frequency $\Lambda$
in the NLS model with Kac-Baker 
($s=0$) long-range interaction
for different values of $\alpha$: $0.1$ 
(full line), $0.3$ (dashed line), $0.7$ (long-dashed line), 
and $\infty$ (dot-dashed line).}
\label{fig:energy-nls}
\end{figure}

\begin{figure}
\includegraphics[width=0.44\textwidth,clip]{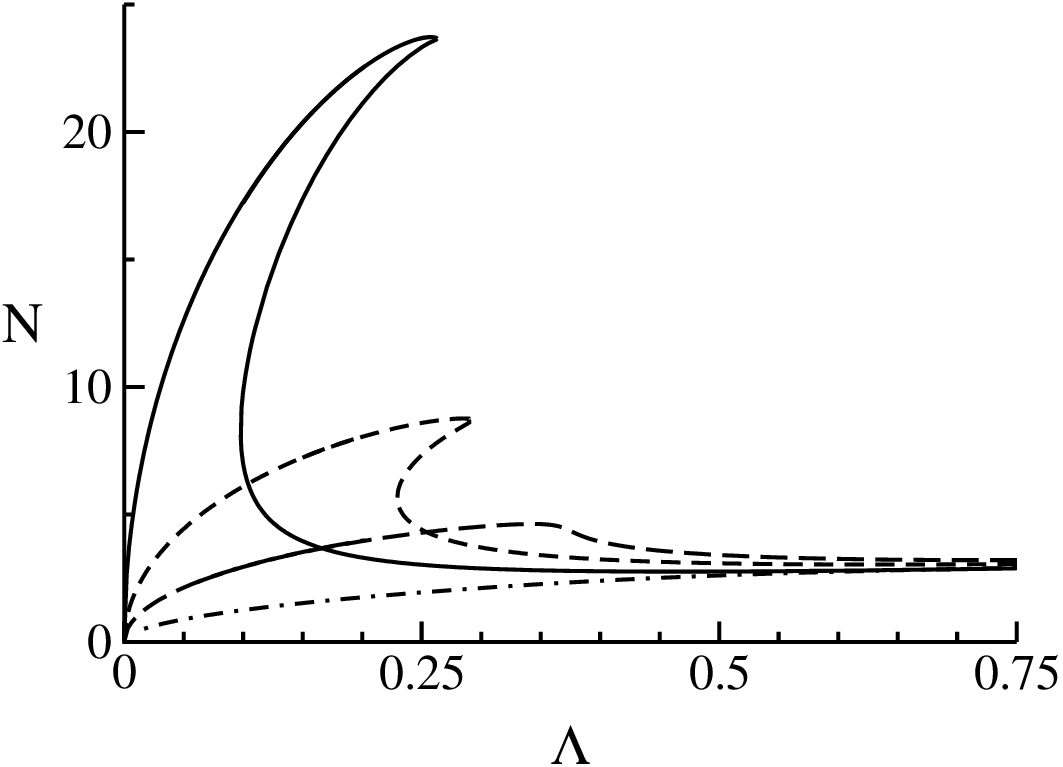}
\caption{Number of quanta $N$ of the
stationary solitons versus nonlinear frequency $\Lambda$
in the NLS model with Kac-Baker 
($s=0$) long-range interaction
for different values of $\alpha$: $0.1$
(full line), $0.3$ (dashed line), $0.7$ (long-dashed line), 
and $\infty$ (dot-dashed line).}
\label{fig:norm-nls}
\end{figure}

\subsection{Nonlinear Schr{\"o}dinger model
with ultra-long-range dispersive interaction}

Let us now consider a system of coupled nonlinear
oscillators which is described by the Hamiltonian
\begin{eqnarray}
\label{ham}
H=\frac{1}{2}\sum_{n} \Biggl\{
\sum_{m \neq n} J_{n-m}(\alpha,s)
|\psi_n-\psi_m|^2 - |\psi_n|^4 \Biggr\} \; ,
\end{eqnarray}
where $n$ and $m$ are site indices, and $\psi_n$ is the
excitation wave function. The first term in Eq. (\ref{ham})
describes a dispersive interaction between nonlinear
oscillators. The second one represents a nonlinear
interaction in the system. The case $\alpha=0$ in the
dispersive interaction has been studied in detail in
Ref. \cite{gmcr}. Here we consider the dispersive
interaction $J_{n-m}(\alpha,s)$ in the Kac-Baker form ($s=0$):
\begin{eqnarray}
\label{KB}
J_{n-m}= J (e^{\alpha}-1)\,e^{-\alpha |n-m|} \; .
\end{eqnarray}
It is important to investigate this type of the dispersive
interaction because real physical systems which are
described by the Hamiltonian (\ref{ham}) with the dispersive
interaction given by  Eq. (\ref{KB}) are not rare. For
example, such type of excitation transfer may take place in
systems where the dispersion curves of two elementary
excitations are close or intersect and effective long-range
transfer occurs at the cost of the coupling between
excitations. This is the case for excitons and photons in
semiconductors and molecular crystals. Dynamics of nonlinear
photonic band-gap materials and periodic nonlinear dielectric
superlattices is described \cite{gcrj} by equations which
are closely related to the equations which govern the
dynamics of the discrete NLS model with the dispersive
interaction in the Kac-Baker form (\ref{KB}).
It was shown in Ref. \cite{gcrj} that the radius of the
effective dispersive interaction is proportional to the
distance between nonlinear layers. Therefore one can easily
tune the range of the dispersive interaction by preparing
nonlinear lattices with different spacing between nonlinear
layers.

In the Hamiltonian (\ref{ham})  $~\psi_n~$ and
$~i \psi^*_n~$ are canonically conjugated variables and
the equation of motion $~i \frac{d}{dt} \psi_n=\partial H /
\partial \psi^*_n~$ has the form
\begin{eqnarray}
\label{deq}
i \frac{d \psi_n}{dt} + \sum_m \, J_{n-m} \,
(\psi_{n-m}-\psi_n) + |\psi_n|^2\psi_n=0 \; .
\end{eqnarray}
The energy of the system $H$ together with the number
of quanta
\begin{eqnarray}
\label{norm}
N=\sum_n\,|\psi_n|^2
\end{eqnarray}
are conserved quantities.

We study stationary states of the system
\begin{eqnarray}
\label{stst}\psi_n(t)=\phi_n(\Lambda)\,e^{i\Lambda t} \; ,
\end{eqnarray}
where $\Lambda$ is the nonlinear frequency and
$\phi_n(\Lambda)$ is the amplitude at the $n$-th site.
For small and intermediate radius ($\alpha^{-1}$) of the
dispersive interaction the problem was studied in Refs.
\cite{jgcr}. It was shown that for $\alpha < 1.7$ the
NLS model exhibits bistability in the spectrum of nonlinear
stationary states: for each value of $N$ in the interval
$[N_l(\alpha),N_u(\alpha)]$
there exist three stationary states with frequencies
$\Lambda_{1}(N) < \Lambda_{2}(N) < \Lambda_{3}(N)$.
Since for discrete NLS equations a necessary and sufficient
condition for linear stability of stationary states with a
single maximum at a lattice site is $d N/d \Lambda\,>\,0$
(see, {\em e.g.}, Ref. \cite{lst}), the low-frequency
$\Lambda_1(N)$ and high-frequency $\Lambda_3(N)$ states,
for which $d N/d \Lambda\,>\,0$, are stable
(in what follows we shall refer to them as to the
`low-frequency branch' and `high-frequency branch').
There is an interval (`gap') of the nonlinear
frequency $\Lambda$ which separates
two branches of nonlinear excitations. Figures
\ref{fig:energy-nls} and \ref{fig:norm-nls} show $H$
versus $\Lambda$ and $N$ versus $\Lambda$, respectively,
obtained from direct numerical solution of Eq. (\ref{deq})
for different values of the inverse radius of dispersive
interaction $\alpha$. It is seen that similar to the case of
the nonlocal Boussinesq equation the gap decreases when the
radius of the dispersive interaction increases and
eventually it vanishes.

\begin{figure}
\includegraphics[width=0.44\textwidth,clip]{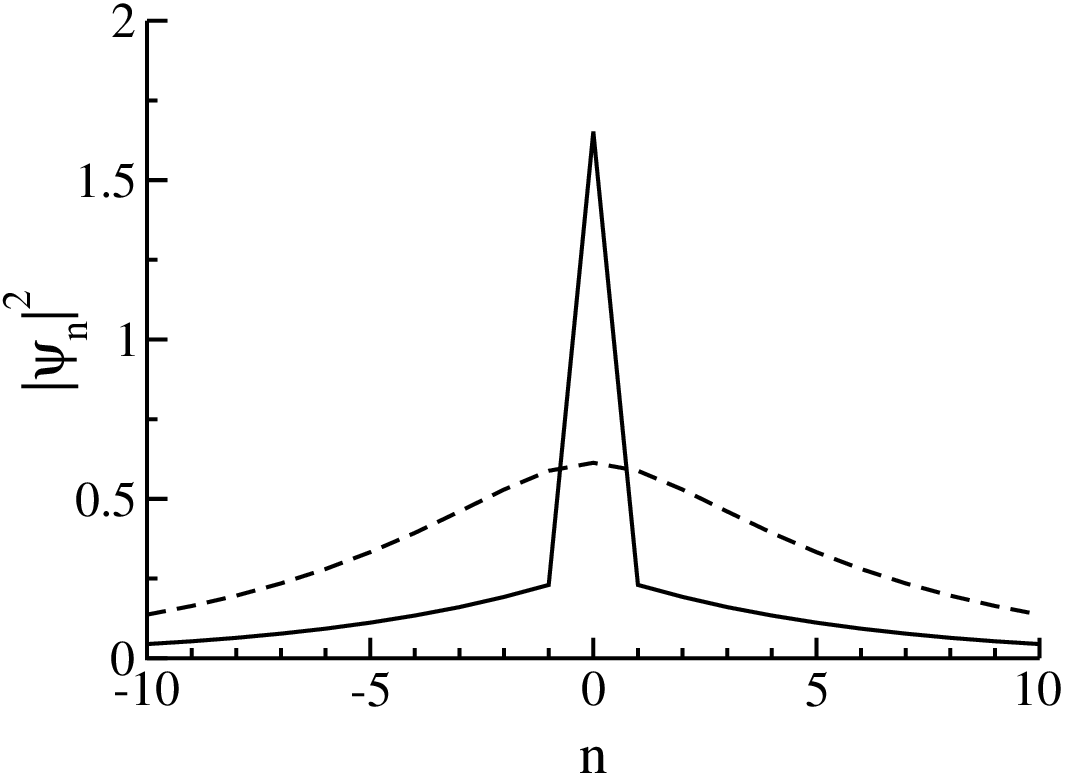}
\caption{Shapes of the high-energy (full line) and
low-energy (dashed line) stationary solitons which coexist
at the same nonlinear frequency $\Lambda=0.25$ in the
NLS model with Kac-Baker long-range interaction for
$\alpha=0.3$ (see Fig. \protect\ref{fig:energy-nls}).}
\label{fig:shape-nls}
\end{figure}

Comparing Figs. \ref{fig:energy:s=0} and
\ref{fig:energy:s=3} obtained for the Boussinesq model
with Fig. \ref{fig:energy-nls} obtained for the NLS model we
see that several new but common for both models features
arise as a consequence of the ultra-long-range dispersive
interaction:
\begin{itemize}
\item The gap which separates two branches of stable
stationary solutions decreases when the radius of the
dispersive interaction $\alpha^{-1}$ increases.

\item For small enough $\alpha$ (in the NLS model for
$\alpha < 0.57$) three different stationary states
coexist for each value of spectral parameter (velocity
$v$ in the case of the Boussinesq equation and nonlinear
frequency $\Lambda$ in the NLS model). These three states
are characterized by different values of the integrals of
motion: the energy $H$ (in both models) and either the
number of quanta $N$ (for the NLS equation) or the
momentum $P$ (for the Boussinesq equation).

\item The shapes of these states differ significantly
(see Figs. \ref{fig:shape:s=0}--\ref{fig:shape:s=3} and
\ref{fig:shape-nls}). The high-energy state is broad and
continu\-um-like while the low-energy states consist of two
components, short-range and long-range ones. The width of
the short-range component is of a few lattice spacing
while  the long-range component has tails which spread out
a few tens of lattice constants.

\item The state with an intermediate energy is always
unstable while the stability of two other states depends
on the radius of the dispersive interaction $\alpha^{-1}$.
For example, in the case of the NLS equation for
$\alpha\,>0.03$ the state which belongs to the
high-frequency branch is unstable ($d N/d \Lambda\,<\,0$
for these $\alpha$). But this state becomes stable when
$\alpha\,<\,0.03$.
\end{itemize}

\section{Analytical approach}

In our analytical approach we will be mostly concerned with
the case of the NLS equation model because an analytical
theory of multi-component solutions of the nonlocal
Boussinesq equation is partially presented in Refs.
\cite{gfnm,mgm}.

\subsection{Nonlinear Schr{\"o}dinger
equation model}

In the quasicontinuum approach, regarding $n$ as a
continuous variable:
$n \rightarrow x,\,\psi_n(t) \rightarrow \psi(x,t)$,
Eq. (\ref{deq}) can be written as
\begin{eqnarray}
\label{peq}
i\partial_t \psi & +& \frac{J}{2} \, Q(\alpha, s=0,
\partial_x) \psi +|\psi|^2 \psi = 0 \; ,
\end{eqnarray}
where
\begin{eqnarray}
\label{pop}
Q(\alpha, s=0, \partial_x) = (e^{\alpha}+1)
\frac{4 \sinh^2(\partial_x/2)}{\kappa^2
     -4 \sinh^2(\partial_x/2)}
\end{eqnarray}
with $\kappa = 2 \sinh(\alpha/2)$ being a range parameter,
is a linear pseudo-differential operator for the Kac-Baker
dispersive interaction (\ref{KB}).

In order to obtain an analytic solution of the
pseudo-differential equation (\ref{peq}) we use an approach
akin to the previously proposed in Ref. \cite{rosenau}: we
multiply Eq. (\ref{peq}) by the operator
$\frac{\partial^2_x}{4\sinh^2(\frac{\partial_x}{2})}$
and expand in a Taylor series this operator
($\frac{\partial^2_x}{4\sinh^2(\frac{\partial_x}{2})}\approx
1-\frac{1}{12} \partial^2_x$)
as well as the operator $\sinh^2(\partial_x/2)$ in the
denominator of the r.h.s. of Eq. (\ref{pop}). We stress that
our modification of the standard continuum approximation
is crucial in the preserving the second length scale that
enters our problem. As a result we get a pseudo-differential
equation
\begin{eqnarray}
\label{nls}
i \Bigl(1-\frac{\partial^2_x}{12} \Bigr) \partial_t \psi
&+& \frac{J}{2}(e^{\alpha}+1) \frac{\partial^2_x}{\kappa^2
-\partial^2_x} \psi + |\psi|^2\psi=0 \; ,
\end{eqnarray}
where the small term $\frac{1}{12}\partial^2_x |\psi|^2\psi$
was omitted.

Looking for stationary solutions
\begin{eqnarray}
\psi=\sqrt{\frac{J}{2}(e^{\alpha}+1)}\,\Phi(x) \,
e^{i\Lambda \tau} \; ,
\end{eqnarray}
where
$\tau=\frac{J}{2}(e^{\alpha}+1) t$,
we can write Eq. (\ref{nls}) in the form
\begin{equation}
\label{eqa}
(\partial^2_x - s^2_+) (\partial^2_x - s^2_-) \Phi
+ \frac{12}{\Lambda} (\partial^2_x - \kappa^2)
\Phi^3=0 \; ,
\end{equation}
with the parameters $s_\pm$ given by
\begin{eqnarray}
\label{spm}
s^2_\pm &=& \frac{1}{2} \Biggl\{ \kappa^2 + 12
\frac{\Lambda+1}{\Lambda} \nonumber \\
&\pm& \sqrt{\left(
\kappa^2 + 12\frac{\Lambda+1}{\Lambda} \right)^2
- 48 \kappa^2} \Biggr\} \; .
\end{eqnarray}
In the limit of small $\alpha$
\begin{eqnarray}
\label{sm}
s^2_- \approx \frac{\Lambda\kappa^2}{\Lambda+1} \; ,
\qquad s^2_+ \approx 12 \frac{\Lambda+1}{\Lambda} \; .
\end{eqnarray}
For small nonlinear frequency $\Lambda$ we expect
excitations with a width much larger than $s^{-1}_{+}$ so
that only one scale $s^{-1}_{-} $ should be important.
Since the characteristic width of the excitations of
interest here is large as compared with $s^{-1}_+$, we
can neglect the operator $\partial^2_x$ in the operator
$(\partial^2_x - s^2_+)$ and  get
\begin{equation}
\label{pl}
(\partial^2_x - s^2_-) \Phi - \frac{1}{\Lambda+1}
(\partial^2_x - \kappa^2)\Phi^3=0 \; .
\end{equation}
This equation was investigated in Ref. \cite{gmck}. It was
shown that Eq. (\ref{pl}) has a soliton solution when
the inequality
$A =  (\kappa^2/s^2_-) \geq 9$
holds. As it is seen from Eq. (\ref{sm}) in the case of
small $\alpha$ this inequality reduces to $\Lambda\,<\,1/8$.
The number of quanta which corresponds to the solution of
Eq. (\ref{pl}) has the form
\begin{eqnarray}
\label{nlf}
N \equiv N_< &=& s_+^2 \Lambda \frac{\kappa}{2 \sqrt{A}}
\Biggl\{ 1 + \frac{A-9}{24} \nonumber \\
&\times& \ln \left( \frac{(\sqrt{A}+1)
(\sqrt{A}+3)}{(\sqrt{A}-1)(\sqrt{A}-3)} \right)
\Biggr\} \; .
\end{eqnarray}
For large $\Lambda$, when both length-scales are important,
we will seek a solution of Eq. (\ref{eqa}) as the sum
\begin{equation}
\label{shl}
\Phi = \sqrt{\frac{\Lambda}{12}} \left\{
\phi_S(x) + \phi_L(x) \right\} \; ,
\end{equation}
where $\phi_S(x)$ is the short-range component and
$\phi_L(x)$ is the long-range one. Here $\phi_S(x)$ will
dominate in the center, while $\phi_L(x)$ in the tails.

Inserting Eq. (\ref{shl}) into Eq. (\ref{eqa}) yields
\begin{eqnarray}
\label{eqsl}
(\partial^2_x -s^2_-) \left\{ (\partial^2_x -s^2_+)
(\phi_S+\phi_L)+(\phi_S+\phi_L)^3 \right\} 
\nonumber \\
-(\kappa^2 - s^2_{-})(\phi_S+\phi_L)^3 =0 \; .
\end{eqnarray}
Assuming that the function $\phi_S(x)$ satisfies the equation
\begin{equation}
\label{eqsh}
(\partial^2_x -s_+^2)\phi_S + 3\phi_L(0)\phi_S^2 +
3\phi_L^2(x)\phi_S+\phi_S^3= 0 \; ,
\end{equation}
we obtain from Eq. (\ref{eqsl}) an equation for $\phi_L(x)$
in the form
\begin{eqnarray}
\label{eql}
(\partial^2_x - s^2_-) (-s_+^2\phi_L+\phi_L^3) +
(s_-^2-\kappa^2) \phi_L^3 \nonumber \\
= B \, (\kappa^2-s_-^2) \, \delta(x) \; ,
\end{eqnarray}
where
\begin{eqnarray}
\label{B}
B &=& \int_{-\infty}^{\infty} dx\, \left\{ \phi^3_S(x) +
3\phi^2_L(x)\phi_S(x) + 3\phi_L(x) \phi^2_S(x) \right\} 
\nonumber \\ 
&=& s_+^2 \int_{-\infty}^{\infty} dx \,\phi_S(x) \; .
\end{eqnarray}
To obtain Eq. (\ref{eql}) we took into account the big
difference in the short-range scale $s_+^{-1}$ and the
long-range scale $s_-^{-1}\,$. It permitted us to replace
the product $~\phi_L(z) \phi_S(z)~$ by
$~\phi_L(0) \phi_S(z)~$ in Eq. (\ref{eqsh}), to neglect
the term $\partial^2_z \phi_L$ as compared with
$s^2_+\phi_L$ in the l.h.s. of Eq. (\ref{eql}) and to
replace the function
$(\kappa^2-s^2_-) \{ \phi^3_S(x)+3\phi^2_L(x)\phi_S(x)+
3\phi_L(x)\phi^2_S(x) \}$
by the r.h.s. of Eq. (\ref{eql}).
Also, Eq. (\ref{eqsh}) was used in Eq. (\ref{B}).

We obtain for the short-range component $\phi_S(x)$ the
expression
\begin{equation}
\label{shsol}
\phi_S(x) =\frac{\sqrt{2} s_+ (1-3 a^2)}
{\sqrt{1-a^2} \cosh(\sqrt{1-3a^2} s_+ x) + \sqrt{2} a} \; ,
\end{equation}
where $a=\phi_L(0)/s_+$ is the amplitude of the long-range
component.

Eq. (\ref{eql}) on the half x-axis subject to the vanishing
at infinity boundary conditions can be presented as
\begin{eqnarray}
\label{eql1}
(s_+^2 &-& 3\,\phi_L^2) \, \partial_x \, \phi_L 
\nonumber \\
&=& - \phi_L \, \sqrt{s_+^4\,s_-^2-s_+^2 \,
\frac{3\,s_-^2 + \kappa^2}{2} \phi_L^2 + \kappa^2
\phi_L^4}
\end{eqnarray}
with the boundary condition at the origin in the form
\begin{eqnarray}
\label{bco}
-2\,\left(s_+^2-3\,\phi^2_L(0)\right)\,\partial_x
\phi_L(+0)&=&(\kappa^2-s_-^2)\,B \; .
\end{eqnarray}
Substituting Eqs. (\ref{B})--(\ref{eql1})
into Eq. (\ref{bco}) we obtain that the amplitude $a$ of the
long-range component is determined by the equation
\begin{eqnarray}
\label{ampl}
\kappa=\frac{a\,R(a)\,A\sqrt{6}}{(A-1) \,
\arccos \sqrt{2a^2/(1-a^2)}} \; ,
\end{eqnarray}
where $R(a)=\sqrt{A\,a^4-(A+3)\,a^2/2+1}\,$.

The number of quanta which corresponds to the solution
(\ref{shl}) is determined by the expression
\begin{eqnarray}
\label{numb}
N \equiv N_> &=& (\Lambda+1) \Biggl( \frac{4}{s_+}
\sqrt{1-3 a^2} + \frac{6}{\kappa \sqrt{A}}
\Biggl\{ 1-R(a) \nonumber \\
&+& \frac{A-9}{12\sqrt{A}} \ln \left|
\frac{4 R(a)+4A a^2-A-3}{4\sqrt{A}-A-3}
\right| \Biggr\} \Biggr) \; .
\end{eqnarray}
When $\kappa\,>\,\kappa_b$, where $\kappa_b \approx0.63$
is the maximum value of the function in the r.h.s. of
Eq. (\ref{ampl}) for $A=9\,(\Lambda=1/8)$,
Eq. (\ref{ampl}) has only one solution in the interval
$A<9\,(\Lambda>1/8)$. Inserting Eq. (\ref{ampl}) into
Eq. (\ref{numb}) and combining Eqs. (\ref{nlf}) and
(\ref{numb}) we obtain that the number of quanta $N$ as
a function of $\Lambda$ has an $N$-type shape
(see Fig. \ref{fig:norm-nls})
with a finite gap which separates two branches of stable
solitons. When $\kappa\,\leq\,\kappa_b$ the gap collapses
and Eq. (\ref{ampl}) has two solutions for
$A>9$ ($\Lambda <1/8$). Taking into account that for
$A>9$ ($\Lambda <1/8$) there exists also a low-frequency
solution of Eq. (\ref{pl}), one can conclude that for
$\kappa\,\leq\,\kappa_b$ there is an interval of
$\Lambda$ where three stationary states correspond to
each value of the nonlinear frequency. In this case the
curve $N(\Lambda)$ has a $Z$-type shape
(see Fig. \ref{fig:norm-nls}).
It is worth noticing that the bifurcation value for the
inverse radius of the dispersive interaction
$\alpha_b\approx 0.61$  ($2\sinh(\alpha_b/2)=\kappa_b$)
obtained analytically is in a surprisingly good agreement
with the value obtained from numerical simulations
($\alpha_b=0.57\pm 0.01$).

It is interesting also to note that when
$\alpha \rightarrow 0$ the amplitude of the long-range
component tends to zero while the short-range component in
the low frequency limit becomes $\delta$-function like.
This result is also in a good qualitative agreement with
the results of numerical simulations
(see Fig. \ref{fig:shape-nls}).

\subsection{ Non-local Boussinesq
equation model}

Here we also restrict ourselves to the case when the
dispersive interaction is of Kac-Baker type. In this case
the quasicontinuum limit of the nonlocal Boussinesq equation
(\ref{sys:eq-st}) can be presented as follows \cite{gfnm}
\begin{equation}
\label{sys:eq-wz}
(\partial_z^2-\sigma_{+}^2) (\partial_z^2-\sigma_{-}^2)
w(z) = \frac{12}{v^2} (\partial_z^2-\kappa^2) w^2(z) \; ,
\end{equation}
where the parameters $\sigma_{\pm}$ play the same role as
the parameters $s_{\pm}$ in the previous subsection.
They are given by
\begin{eqnarray}
\label{sys:spm}
\sigma^2_\pm &=& \frac{1}{2} \, \Biggl\{ \kappa^2 +
12 \frac{v^2-1}{v^2} \nonumber \\
&\pm& \, \sqrt{\left( \kappa^2 -
12 \frac{v^2-1}{v^2} \right)^2 + 48 \kappa^2
\frac{c^2-1}{v^2}} \Biggr\} \; ,
\end{eqnarray}
where  the speed of sound $c$ in the lattice with the
Kac-Baker interaction is determined by the expression
\begin{equation}
\label{speed}
c = \left[ 1 +  J \frac{{\rm e}^{-\alpha} + 1}{
({\rm e}^{-\alpha} - 1)^2} \right]^{1/2} .
\end{equation}
The parameter $\sigma_{+}$ is finite at all velocities
$v \geq c$ and tends to $\sqrt{12}$ for $v \to \infty$.
The parameter $\sigma_{-}$ vanishes at $v=c$ and tends to
$\kappa$ for $v \to \infty$.
One can  find that in the travel-wave approach the Hamiltonian
is determined by the expressions
\begin{eqnarray}
\label{hamb}
H=v P - M(v) \; ,
\end{eqnarray}
where $P$ is the canonical momentum and
\begin{eqnarray}
\label{nl}
M(v)=\frac{1}{6}\int_{-\infty}^{\infty} \, w^3(x) d x
\end{eqnarray}
is the nonlinear part of the Hamiltonian.

Near the speed of sound excitations have a width much
larger than $\sigma^{-1}_{+}$ and can be described by
Eq. (\ref{sys:eq-wz}) in the approximation
$(\partial_z^2 - \sigma_{+}^2) w \approx - \sigma_{+}^2 w$.
In this approximation the soliton solutions exist in a
finite interval of velocities,
$c<v<v_{cr} \simeq \sqrt{(4c^2-1)/3}\,$.
The explicit form of the low-velocity soliton in given in
Ref. \cite{gfnm}. Using this expression it is
straightforward to obtain the energy of the soliton
belonging to the low-velocity branch in the form
\begin{eqnarray}
\label{hlvb}
H_<=v P_ < - M_<(v) \; , \nonumber\\
P_<=\frac{ \sigma_+^4 v^5}{72 {\cal A}^2 \sigma_-}
\left(\frac{{\cal A}-4}{3} I(w_1)+2 \right) \; , \nonumber\\
M_<(v)=\frac{\sigma_+^6 v^6 }{93312 \,{\cal A}^3}
\biggl({\cal A}^2-2 {\cal A} + 40 \nonumber\\
- \frac{({\cal A}-4)^2 ({\cal A}+5)}{3} I(w_1) \biggr) \; ,
\end{eqnarray}
with
\begin{eqnarray}
\label{int}
I(w) &=& \int\limits_{0}^{w}\frac{1}{{\cal R}(x)}dx 
\nonumber\\
&=& \frac{1}{\sqrt{{\cal A}}}\,\ln \left( \left|
\frac{\sqrt{{\cal A} {\cal R}(w)} + {\cal A} w -
({\cal A}+2)/3}{\sqrt{{\cal A}}-({\cal A}+2)/3}
\right| \right) \; , \nonumber\\
{\cal R}(x) &=& \sqrt{{\cal A} x^2 - \frac{2}{3}
({\cal A}+2) x + 1} \; ,
\end{eqnarray}
where
${\cal A} \equiv (\kappa^2/\sigma_-^2) > 4$
in the velocity interval $c<v<v_{cr}$, and
\begin{eqnarray}
w_1 = \frac{1}{3 {\cal A}} \left( {\cal A}+2 -
\sqrt{({\cal A}-1)({\cal A}-4)} \right)
\end{eqnarray}
is the amplitude of the low-velocity soliton.

Far from the speed of sound we will use the same approach
as in the previous subsection (see also Ref. \cite{gfnm}).
We seek a solution of Eq. (\ref{sys:eq-st}) as a sum
\begin{equation}
\label{wsl}
w(z) = w_S(z) + w_L(z) \; ,
\end{equation}
where $w_S(z)$ is the short-range component of the strain
and $w_L(z)$ is the long-range one.

Assuming that the function $w_S(z)$ satisfies the equation
\begin{equation}
\label{eqws}
\partial^2_z w_S - \left( \sigma^2_+
+ \frac{24}{v^2 }w_L(0) \right) w_S
- \frac{12}{v^2} w^2_S = 0 \; ,
\end{equation}
we obtain from Eq. (\ref{sys:eq-wz}) an equation for
$w_L(z)$ in the form
\begin{eqnarray}
\label{eqwl}
\partial^2_z (w_L+\frac{12}{v^2 \sigma^2_+} w^2_L) -
\sigma_-^2 w_L-\frac{12 \kappa^2}{v^2\,\sigma_+^2} \,
w^2_L \nonumber\\
= - {\cal B} \, (\kappa^2 - \sigma^2_{-}) \, 
\delta(z) \; ,
\end{eqnarray}
where
\begin{equation}
\label{cB}
{\cal B} = \int_{-\infty}^{\infty} dz w_S(z) \; .
\end{equation}
To obtain Eqs. (\ref{eqws}) and (\ref{eqwl}) we took into
account the big difference in the short-range scale
$\sigma_+^{-1}$ and the long-range one
$\sigma_-^{-1}$ and proceeded in the same way as we did in
the previous subsection. From Eq. (\ref{eqws}) we get the
short-range component $w_S$ in the form
\begin{equation}
\label{wsr}
w_S(z) = - \frac{1}{4} \sigma^2_+ v^2 (1-2\gamma) \,
\sech^2 \left(\sqrt{1-2\gamma}\,\frac{\sigma_+ z}{2}
\right) \; ,
\end{equation}
where $\gamma=12 \, |w_L(0)| \,/\, (v \sigma_+)^2$
determines the amplitude of the long-range component.
It can be found from the equation \cite{gfnm}
\begin{eqnarray}
\label{eqgam}
3 ({\cal A}-1)\,\frac{\sigma_-}{\sigma_+} =
\frac{\gamma {\cal R}(\gamma)}{\sqrt{1-2 \gamma}} \; .
\end{eqnarray}
Note that there is a misprint in the corresponding equation
of Ref. \cite{gfnm}: one should omit $\bar{w}_L(0)$
(which plays the same role as $\gamma$ in this paper) in
the nominator of the fraction in the l.h.s. of Eq. (A.19).

Using Eqs. (\ref{wsr}), (\ref{wsl}), and (\ref{eqgam}) one
can obtain that the energy of the compound soliton
(\ref{wsl}), {\em i.e.} the soliton which belongs to the
high-velocity branch, can be presented in the form
\begin{eqnarray}
\label{hhvb}
H_>=v P_> - M_>(v) \; ,
\end{eqnarray}
where
\begin{eqnarray}
\label{mhvb}
P_> &=& \frac{\sigma_+^4 v^5}{72}
\biggl( \frac{3}{\sigma_+}\sqrt{1-2 \gamma} 
\nonumber\\
&+& \frac{1}{{\cal A}^2 \sigma_-} \left[ 2+ 
\frac{{\cal A}-4}{3}
I(\gamma)-(2+{\cal A} \gamma){\cal R}(\gamma)
\right] \biggr)
\end{eqnarray}
is the canonical momentum of the compound soliton and
\begin{eqnarray}
\label{nl2}
M_>(v)=\frac{1}{2880} \sigma_+^5 v^6  \sqrt{1-2 \gamma}
\left(3 \gamma^2-3 \gamma+2 \right) \nonumber\\
+ \frac{\sigma_+^6 v^6 }{93320 {\cal A}^3}
\biggl( ({\cal A}^2-2 {\cal A}+40) (1-{\cal R}(\gamma)) 
\nonumber\\
- {\cal A}({\cal A}+20)\gamma {\cal R}(\gamma)-
\frac{({\cal A}-4)^2 ({\cal A}+5)}{3} I(\gamma) \biggr)
\end{eqnarray}
is the corresponding nonlinear part.

\begin{figure}
\includegraphics[width=0.44\textwidth,clip]{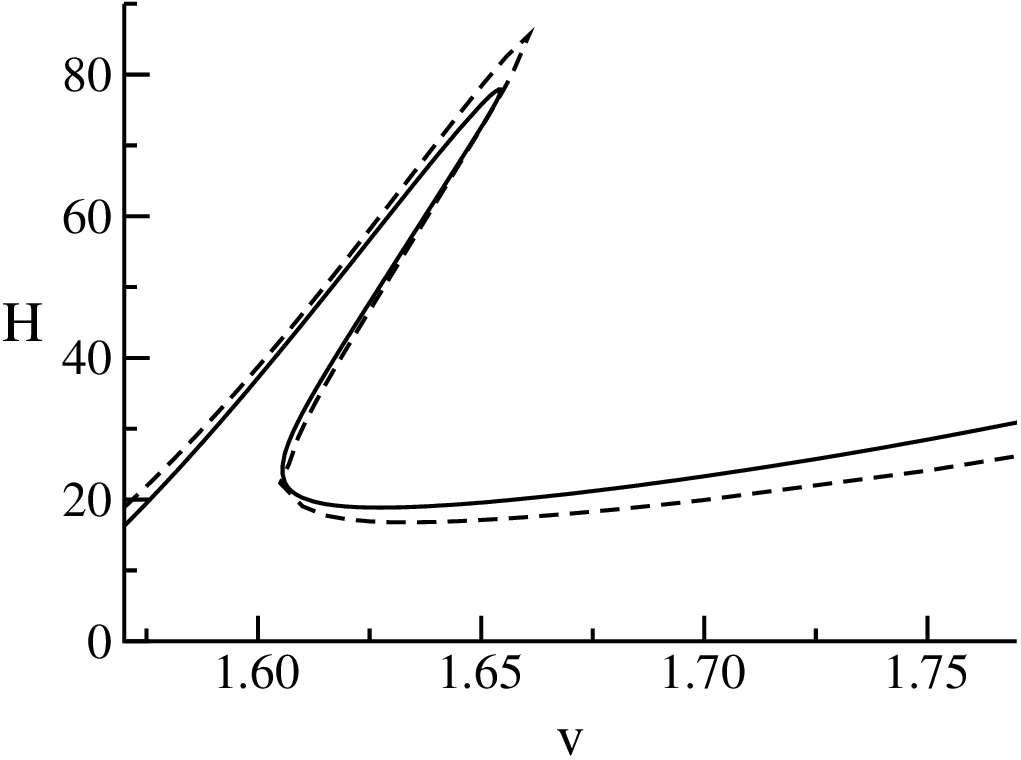}
\caption{Energy of the pulse solitons versus velocity
in the anharmonic lattice with Kac-Baker 
($s=0$) long-range
interaction calculated numerically (full line) and
analytically (dashed line) for $\alpha=0.05$ and
$J=0.0016$.}
\label{fig:anal:s=0}
\end{figure}

In the velocity interval $v\,>\,v_c$ (${\cal A}\,<\,4$)
Eq. (\ref{eqgam}) for all values of the inverse radius
$\alpha$  has a solution which tends to zero when
$v \rightarrow \infty$. There is an one-to-one
correspondence between the velocity of soliton and its
energy. But when
\begin{eqnarray}
\label{mult}
\frac{\kappa^2 (4 c^2-1)}{c^2-1}\leq \frac{64}{729}
\end{eqnarray}
Eq. (\ref{eqgam}) has also two solutions in the interval
$v\,<\,v_c ~({\cal A}\,>\,4)~$. In other words under the
condition (\ref{mult}), which taking into account the
explicit form of the speed of sound (\ref{speed}),
can be written as
\begin{eqnarray}
\label{multi}
J\geq \frac{3 \alpha^4}{8 (\alpha_o^2-\alpha^2)}
\end{eqnarray}
with $\alpha_o\approx 0.16$, there exist three stationary
pulse solitons with different energy for each value of the
velocity and the function $H(v)$ has a characteristic
$Z$-shape (see Fig. \ref{fig:energy:s=0}).
Comparison the dependence $H(v)$ obtained analytically
from Eqs. (\ref{hlvb}), (\ref{hhvb})-(\ref{nl2}) with the
dependence obtained numerically shows a good agreement
between these two approaches (see Fig. \ref{fig:anal:s=0}).
The function (\ref{multi}) also describes very well the
boundary which separates the area where $H(v)$ has a
$N$-shape and a $Z$-shape (see Fig. 3 in Ref. \cite{mgm}).

\section{ Conclusion}

We investigated the effect of length scale competition in
two systems which can bear pulse-like nonlinear excitations:
a chain of coupled nonlinear oscillators (nonlinear
Schr{\"o}dinger model) and an anharmonic chain with cubic
nearest-neighbor interaction. Two types of dispersive
interactions were studied: the Kac-Baker interaction and
screened Coulomb interaction. We have shown that the
existence of two competing length scales: the radius of the
dispersive interaction and lattice spacing provides the
existence of two types of pulse solitons one of which is
broad, continuum-like and uni-component, and another is
narrow, discrete-like and has a compound structure.
It consists of two components: short-ranged breather-like
component which dominates near the center of the excitation
and long-ranged component which contributes mainly to the
tails of the excitation. These two stable solitons are
separated by gap: an interval of spectral parameter
(velocity in the case of Boussinesq equation and frequency
in the case of Nonlinear Schr{\"o}dinger equation) where
there are no stable stationary excitations. We show that
the gap decreases when the  radius of the dispersive
interaction increases and it closes when the radius of the
dispersive interaction exceeds some critical value.
In contrast to systems with short-range and moderate-range
dispersive interactions, where there is unambiguous
correspondence between the velocity (frequency in the NLS
equation) of the soliton and its energy, in systems where
the length scales differ significantly (ultra-long-range
interaction) there is an interval where three different
states with different energies (momentum, number of quanta,
etc.) correspond to each value of the spectral parameter.
This is a generic property of all systems with competing
length scales and is due to the compound structure of
excitations.

\section*{Acknowledgements}

Two of the authors (Y. G. and S. M.) are grateful for the
hospitality of the University of Bayreuth and the Technical
University of Denmark (Lyngby) where this work was done.
We also acknowledge the support provided by the DRL Project
No. UKR-002-99. S. M. acknowledges support from the 
European Commission RTN project LOCNET 
(HPRN-CT-1999-00163).




\begin{thebibliography}{99}

\bibitem{flach}
S. Flach and C.R. Willis, Phys. Rep. {\bf 295}, 181 (1998).

\bibitem{lst}
E.W. Laedke, K.H. Spatschek, and S.K. Turitsyn,
Phys. Rev. Lett. {\bf 73}, 1055 (1994);
E.W. Laedke, O. Kluth,  K.H. Spatschek,
Phys. Rev. E {\bf 54}, 4299 (1996).

\bibitem{mal}
B.A. Malomed and M.I. Weinstein, Phys. Lett. A
{\bf 220}, 91 (1996).

\bibitem{kladko}
S. Flach, K. Kladko, and R.S. MacKay,
Phys. Rev. Lett. {\bf 78}, 1207 (1997).

\bibitem{ngfm}
A. Neuper, Yu. Gaididei, N. Flytzanis, and F.G. Mertens,
Phys. Lett. A {\bf 190}, 165 (1994).

\bibitem{gfnm}
Yu. Gaididei, N. Flytzanis, A. Neuper, and F.G. Mertens,
Phys. Rev. Lett. {\bf 75}, 2240 (1995);
Physica D {\bf 107}, 83 (1997).

\bibitem{mgm}
S.F. Mingaleev, Yu.B. Gaididei, and F.G. Mertens,
Phys. Rev. E {\bf 58}, 3833 (1998).

\bibitem{gmck}
Yu.B. Gaididei, S.F. Mingaleev, P.L. Christiansen, and
K.{\O}. Rasmussen, Phys. Lett. A {\bf 222}, 152 (1996).

\bibitem{gmcr}
Yu.B. Gaididei, S.F. Mingaleev, P.L. Christiansen, and
K.{\O}. Rasmussen, Phys. Rev. E {\bf 55}, 6141 (1997).

\bibitem{mks}
S.F. Mingaleev, Yu.S. Kivshar, and R.A. Sammut,
Phys. Rev. E {\bf 62}, 5777 (2000).

\bibitem{jgcr}
M. Johansson, Yu. B. Gaididei, P.L. Christiansen, and
K.{\O}. Rasmussen, Phys. Rev. E {\bf 57}, 4739 (1998).

\bibitem{jagcr}
M. Johansson, S. Aubry, Yu.B. Gaididei, P.L. Christiansen,
and K.{\O}. Rasmussen, Physica D {\bf 119}, 115 (1998).

\bibitem{gcrj}
Yu.B. Gaididei, P.L. Christiansen, K.{\O}. Rasmussen, and
M. Johansson,  Phys. Rev. B {\bf 55}, R13365 (1997).

\bibitem{kov}
M.V. Gvozdikova and A.S. Kovalev, Low Temp. Phys.
{\bf 24}, 808 (1998) [Fiz. Nizk. Temp. {\bf 24}, 1077 (1998)].

\bibitem{mgm00}
S.F. Mingaleev, Yu.B. Gaididei, and F.G. Mertens,
Phys. Rev. E {\bf 61}, R1044 (2000).

\bibitem{htf53} W. Magnus, F. Oberhettinger and
R.P. Soni, {\em Formulas and Theorems for the
Special Functions of Mathematical Physics}
(Springer-Verlag, Berlin -- Heidelberg -- New York,
1966).

\bibitem{rosenau}
P. Rosenau, Phys. Lett. A {\bf 118}, 222 (1986).

\end{thebibliography}
\end{document}